\documentclass[aps,prd,nofootinbib,twocolumn,10pt,floatfix]{revtex4-1}
\pdfoutput=1
\usepackage[utf8]{inputenc}
\usepackage[T1]{fontenc}
\usepackage{amssymb}
\usepackage{amsfonts}
\usepackage[intlimits]{mathtools}
\usepackage[english]{babel}
\usepackage{lmodern}
\usepackage{graphicx}
\usepackage{ifpdf}
\usepackage{nicefrac}
\usepackage{calc}

\ifpdf
\else
  \PassOptionsToPackage{dvipdfmx}{hyperref}
\fi

\usepackage{hyperref}

\usepackage{bbm}
\pdfpkmode={supre}
\newcommand{\bbs}{\mathbbm{s}}
\newcommand{\bbp}{\mathbbm{p}}
\newcommand{\bbv}{\mathbbm{v}}
\newcommand{\bba}{\mathbbm{a}}
\newcommand{\bbt}{\mathbbm{t}}
\newcommand{\bbw}{\mathbbm{w}}

\newcommand{\vac}{\mathrm{vac.}}
\newcommand{\Eqref}[1]{Eq.~\eqref{#1}}
\newcommand{\Figref}[1]{Fig.~\ref{#1}}
\DeclareMathOperator\tr{tr}

\newcommand{\e}{{\text{e}}}
\newcommand{\cm}{{q}}
\renewcommand{\i}{{\text{i}}}
\newcommand{\ii}{{\text{i}}}
\renewcommand{\Re}{{\text{Re}}}
\renewcommand{\Im}{{\text{Im}}}

\newcommand{\Eqs}{{Eqs.~}}
\DeclareMathOperator\Rect{Rect}
\DeclareMathOperator\E{\text{\textbf{E}}}
\DeclareMathOperator\K{\text{\textbf{K}}}
\DeclareMathOperator\PiI{\boldsymbol{\Pi}}
\DeclareMathOperator\artanh{artanh}

\newlength{\colonwidth}
\newcommand{\definedby}{\hspace*{\colonwidth} &\mathrel{\hspace*{-\colonwidth}:=}}

\newcommand{\commutator}[2]{\left[ #1 ,\, #2 \right]}

\newcommand{\todo}[1]{\marginpar{\textit{todo}}}

\let\oldvec\vec
\renewcommand{\vec}[1]{\hspace{-0.3pt}\oldvec{\hspace{0.3pt}#1}}

\let\oldwidehat\widehat
\renewcommand{\widehat}[1]{\hspace{-0.2pt}\oldwidehat{\hspace{0.2pt}#1}}

\begin{document}
%-------------------------------------------------------------------------------

\hypersetup{pdftitle={Comparison of semiclassical and Wigner function methods in pair production in rotating fields},
            pdfauthor={Alexander Blinne, Eckhard Strobel}}
\title{Comparison of semiclassical and Wigner function methods in pair production in rotating fields}
\author{Alexander Blinne}
\email{alexander.blinne@uni-jena.de}
\affiliation{Theoretisch-Physikalisches Institut,  Abbe Center of Photonics}
\affiliation{Friedrich-Schiller-Universität Jena, Max-Wien-Platz 1, D-07743 Jena, Germany}
\affiliation{Helmholtz-Institut Jena, Fröbelstieg 3, D-07743 Jena, Germany}
\author{Eckhard Strobel}
\affiliation{ICRANet, Piazzale della Repubblica 10, 65122 Pescara, Italy}
\affiliation{Dipartimento di Fisica, Universit\`a di Roma "La Sapienza", Piazzale Aldo Moro 5, 00185 Rome, Italy}
\affiliation{Universit\'e de Nice Sophia Antipolis, 28 Avenue de Valrose, 06103 Nice Cedex 2, France}

\begin{abstract}
We present a comparison of two methods to compute the momentum spectrum and the Schwinger pair creation rate for pulsed rotating electric fields: 
one based upon the real-time Dirac--Heisenberg--Wigner (DHW) formalism and a semiclassical approximation based on a scattering ansatz.
For the semiclassical method we propose to either perform numerical calculations or an additional approximation based on an analytical solution for the constant rotating field.
We find that both numerical methods are complementary with respect to computation time as well as accuracy. 
The approximate method shows the same qualitative features while being computationally much faster. 
We additionally find that the unequal production of pairs in different spin states reported for constant rotating fields with the scattering method is in agreement with the Wigner function method.
\end{abstract}

\maketitle

%-------------------------------------------------------------------------------
\section{Introduction}
%-------------------------------------------------------------------------------
The concept of a strong electric field producing electron positron pairs from the vacuum, usually called Schwinger pair production, was first discussed by Sauter in 1931 \cite{Sauter1931}.
Since then it has been studied in detail in more complicated field configurations. 

Different methods were developed including those that are exact on the mean-field level e.\,g. the quantum kinetic theory (QKT) \cite{Schmidt1998} and the real-time Dirac--Heisenberg--Wigner (DHW) formalism \cite{Bialynicki-Birula1991,Hebenstreit2011}.
Using a scattering ansatz it is still possible to find exact results, but the ansatz is often combined with a semiclassical approximation \cite{Brezin1970,Popov1971,Popov1972,Popov1973,Marinov1977,Kim:2000un,Popov2001,Kim:2003qp,Piazza:2004sv,Kim:2007pm,Kleinert2008,Kleinert2013}.
This combination is sometimes referred to as Wentzel--Kramers--Brillouin (WKB) approach \cite{Dumlu2011,Strobel2014,Dumlu2015,Ilderton2015Interpolating} or even as WKB approximation \cite{Strobel2014,Dumlu2009}, while only the ansatz, but not the approximation, is taken from the original WKB method.
We will thus not refer to the method discussed here as WKB method, but as semiclassical scattering method.
Other semiclassical methods include the worldline instanton method \cite{Dunne2005B,Dunne2006}.

In semiclassical calculations simple field configurations result in pair creation rates which are the same up to a factor of 2  in scalar and spinor QED respectively.
However if there is more than one pair of classical turning points, interference effects arise which lead to differing results \cite{Dumlu2010,Dumlu2011,Dumlu2011WLI}.

The first extensions of Sauter's original work concentrated on one-component electric fields.
In addition to one-component fields depending on either space \cite{Nikishov:2003ig} or time, exact solutions can be found in lightcone variables \cite{Tomaras2001,Hebenstreit2011B}. 
In \cite{Ilderton2015Interpolating} a connection between these three special cases was found using interpolating coordinates and the worldline instanton method.

Recently also more involved fields have been studied including electric fields with two or three components depending on space \cite{Dunne2006B} and fields depending on space as well as on time \cite{Schneider2014}.
In the presence of very strong fields QED cascades of successive radiation of accelerated charges and particle production from hard photons are expected to occur \cite{Bell2008, Kirk2009, Fedotov2010, Bulanov2010B, Elkina2011, Nerush2011, Nerush2011B, King2013, Bulanov2013, Bashmakov2014} and a lot of the research in this area concentrates on rotating fields.
A lot of interest is put upon field configurations that could be found in counterpropagating lasers, this includes (nonlinear) Breit--Wheeler pair production \cite{Nousch:2012hg,Nousch:2015pja} and pair production in pure electric fields near the antinodes of the magnetic field \cite{Hebenstreit:2010cc,Jansen:2013dea,Akal:2014eua,Otto:2015gla,Panferov:2015yda}.

The leading semiclassical order (often referred to as exponential factor) of Schwinger pair production for constant rotating fields has been studied using the scattering method in \cite{Popov1973,Strobel2014} and using the worldline instanton method in \cite{Xie2012}.
This was extended to include the next order (often referred to as prefactor) in \cite{Strobel2015} for the scattering method and in \cite{Ilderton2015Interpolating} in interpolating coordinates for the worldline instanton method.
Pair production for a pulsed rotating field has first been studied in \cite{Blinne2013} using a method based upon the Dirac--Heisenberg--Wigner formalism, which we will call the Wigner function method or, in short, the Wigner method.
For generalizations to elliptically polarized fields see \cite{Li:2014nua}.

In the current work we compare the results of the semiclassical scattering method and the Wigner method of \cite{Strobel2015} and \cite{Blinne2013} respectively for rotating electric field pulses of the form
\begin{equation}
  \label{eqn:puls-sauter-rot}
  \vec{E}(t)=E(t)\begin{pmatrix}
                    \cos(\Omega t) \\
                    \sin(\Omega t) \\
                    0
                  \end{pmatrix}\,,
\end{equation}
where \(E(t)\) defines the shape of the pulse. 
For a fair comparison, we mostly stay in the semiclassical regime.
In contrast to the case of a constant rotating pulse, it is difficult to find an analytical solution for general $E(t)$ in the semiclassical scattering approach.
We thus propose two different ways to proceed.
The first is to carry out the necessary computation numerically, the second is to perform an additional approximation which uses the analytic results for the constant rotating field. 
We will refer to the latter one as locally constant rotating field approximation (LCRFA).

Comparing the numerical methods we find that they are complementary with respect to computation time as well as with respect to computational accuracy.
The LCRFA is computationally much faster, behaves qualitatively like the numerical methods and is a good approximation for long enough pulses.
Additionally we find that the two independent solutions found for the scattering method can be connected to a linear combination of spinor states with specific chirality and magnetic moment.
With this knowledge we succeed in reproducing these two independent solutions using the Wigner method.

This paper is structured as follows: 
in the first three sections we briefly introduce the methods which are compared later.
In section \ref{sec:wigner} we review the Wigner method and present a method to project the Wigner function onto specific spinor states.
The results of the scattering method for two-component electric fields are derived directly from the Dirac equation in section \ref{sec:semiclass}.
The new LCRFA for rotating electric fields is introduced in section \ref{sec:Approx}.
In section \ref{sec:cmp} we compare the three methods with respect to accuracy and computation time and we summarize our conclusions in section \ref{sec:concl}.

To make the main ideas more transparent we displace the more technical discussion of the numerical methods to the appendices \ref{sec:numwigner} and \ref{sec:numsemi}. For self-containedness the analytic results for the constant rotating pulse and the non-rotating Sauter pulse can be found in appendices \ref{app:constrot} and \ref{app:Sauter} respectively.

%-------------------------------------------------------------------------------
\section{The Wigner function}\label{sec:wigner}
%-------------------------------------------------------------------------------

The equal-time Wigner function $\mathcal{W}$ is defined as the vacuum expectation value of the Wigner operator $\mathcal{\widehat{W}}$ given by
\begin{align*}
\begin{split}
  \mathcal{\widehat{W}}_{ab}(\vec{x},\vec{p},t)\definedby
  -\frac12\int d\vec{s}\,\e^{-\ii\vec{p}\cdot\vec{s}}
\, \e^{-ie\int_{\vec{x}+\nicefrac{\vec{s}}{2}}^{\vec{x}-\nicefrac{\vec{s}}{2}}\vec{\hat{A}}(t,\vec{x}'\!)\cdot d\vec{x}'}
\\
&\qquad\qquad \cdot \commutator{\hat{\Psi}_a(t,\vec{x}+\nicefrac{\vec{s}}{2})}{\hat{\overline{\Psi}}_b(t,\vec{x}-\nicefrac{\vec{s}}{2})}
\end{split}
\\
\mathcal{W}\definedby \langle0|\mathcal{\widehat{W}}|0\rangle\,.
\end{align*}
For details of the formalism we refer to \cite{Bialynicki-Birula1991, Hebenstreit2010, Blinne2013}.
In general, the equal-time Wigner function of the Dirac field can be expressed by its components according to the Fierz decomposition
\begin{align}
 \label{eqn:wigner_fierz}
  \mathcal{W}(\vec{p},\vec{x},t) = \frac14
      (
        \mathbbm{1} \bbs + \ii \gamma_5 \, \bbp
        +\gamma^\mu \, \bbv_\mu + \gamma^\mu \gamma_5 \, \bba_\mu
        +\sigma^{\mu\nu} \, \bbt_{\mu\nu}
      )\,.
\end{align}
In total these are 16 independent real components.
The Wigner function for a pure vacuum can be calculated directly from the definition \cite{Bialynicki-Birula1991} and only four components are nonzero
\begin{align}
  \label{eqn:wigner_vac}
    \quad \bbs_\vac
  &=  \frac{-2m}{\omega}
  \,,\quad \vec{\bbv}_\vac
  =   \frac{-2\vec{p}}{\omega}\,,
\end{align}
where
\begin{align}
 \omega^2\definedby \vec{p}^2+m^2.  \label{eq:scE}
\end{align}
In general the components satisfy a system of coupled PDEs that follows from the Heisenberg equation of motion for the fermionic field operators.
Anywhere but in the Wilson line the electromagnetic field is purely treated on the mean field level $\hat{\vec{A}}\to\vec{A}$. 
In a spatially homogeneous setup at most ten of the 16 components are nonzero, specifically
\begin{align*}
  \bbw&=\begin{pmatrix}
    \bbs, & \vec{\bbv}, & \vec{\bba}, & \vec{\bbt}\,
  \end{pmatrix}^\intercal\,, &
  \left(\vec{\bbt}\right)_i \definedby  \bbt_{0i}-\bbt_{i0}\,.
\end{align*}

The one particle distribution function $f$ can be calculated from the phase space energy density
\begin{align*}
  \varepsilon &= m\bbs+\vec{p}\cdot\vec{\bbv}
\end{align*}
by normalizing to the energy of one particle pair after subtracting the vacuum solution, thus
\begin{align*}
  f &= \frac{1}{2\omega}\left( \varepsilon - \varepsilon_\mathrm{vac.} \right)\,.
\end{align*}
This formula can be written in terms of a projection of the Wigner function
\begin{align}
  \label{eqn:general_f}
  f[\mathcal{W} - \mathcal{W}_\mathrm{vac.}] &= \frac{1}{2\omega}\tr\left[
	  \left( \mathcal{W} - \mathcal{W}_\mathrm{vac.} \right)
	  \left( m\mathbbm{1}+\vec{p}\cdot\vec{\gamma} \right)
	  \right]\,.
\end{align}

Starting from these definitions and using the Dirac equation, a modified quantum kinetic equation \cite{Blinne2013} can be derived, which can be solved numerically to calculate the one particle distribution function $f(t,\vec{p})$ at $t\to\infty$.
This involves the method of characteristics, which transforms the partial differential equations into ordinary ones by requiring that the kinetic momentum $\vec{p}$ follows the solution of the classical equation of motion for a positron in the external field with canonical momentum $\vec{q}$
\begin{align*}
  \vec{p}(t) &=\vec{q} -e\vec{A}(t) \,.
\end{align*}
Since results are always being taken for $t\to\infty$ it is worthwhile to gauge the vector potential $\vec{A}$ in such a way that $\vec{p}(t)\to\vec{q}$ for $t\to\infty$\,.

The modified quantum kinetic equation is a system of 10 equations for $f$ and 9 auxiliary quantities $\vec{v}, \vec{a}, \vec{t}$ which can be identified with the aforementioned components according to
\begin{align}
  \label{eqn:wigner_subst}
  \begin{aligned}
 \bbs(\vec{p}(t), t)       &= (1-f(\vec{q}, t))\, \bbs_\vac(\vec{p}(t), t) - \vec{p}(t) \cdot \vec{v}(\vec{q}, t) \,,\\
 \vec{\bbv}(\vec{p}(t), t) &= (1-f(\vec{q}, t))\, \vec{\bbv}_\vac(\vec{p}(t), t) + \vec{v}(\vec{q}, t) \,,\\
 \vec{\bba}(\vec{p}(t), t) &= \vec{a}(\vec{q},t) \,,\\
 \vec{\bbt}(\vec{p}(t), t) &= \vec{t}(\vec{q},t)\,.
 \end{aligned}
\end{align}
From now on $\vec{p}(t)$, which clearly also depends parametrically on $\vec{q}$ will be denoted by just $\vec{p}$ for the remainder of this section.
The electric field $\vec{E}$ is given by $\vec{E}(t)=-\dot{\vec{A}}(t)$ and the modified quantum kinetic equations read
\begin{align*}
       \dot{f}
      &=\frac{e}{2\omega}\vec{E}\cdot\vec{v} \,,
    \\
    \begin{split}
    \dot{\!\oldvec{\,v}}
      &=\frac{e}{2\omega^3}\left( \vec{p} (\vec{E}\cdot\vec{p})-\omega^2\vec{E} \right)(f-1)\\ &\hspace{4cm}
      -\frac{e}{\omega^2}\vec{p}(\vec{E}\cdot\vec{v})
     -\vec{p}\times\vec{a}
      -2\vec{t} \,,
    \end{split}
    \\
    \,\dot{\!\vec{a}}
      &=-\vec{p}\times\vec{v} \,,
    \\
    \dot{\!\vec{\,t}}
      &=2\left(\vec{v}+\vec{p}(\vec{p}\cdot\vec{v})\right)\,.
\end{align*}
Combined with the initial condition
\[
f = 0,\ \vec{v}=\vec{a}=\vec{t}=\vec{0}
\]
at $t\to-\infty$\,, the initial value problem is well defined.

In this work the numerical integration has been carried out using the Runge--Kutta--Cash--Karp--54 scheme as implemented as part of the C++ library \textit{Boost.Numeric.Odeint} \cite{Ahnert2011}.
In order to sample the momentum distribution, a grid of values for the canonical momentum $\vec{q}$ is chosen and the initial value problem is solved for each grid point.
As the calculations for each grid point are independent, the computation is easily parallelized.
The results shown in this work have been calculated using the \textit{Omega}-Cluster at \textit{FSU Jena}.
A few more details about the numerical calculations are explained in appendix \ref{sec:numwigner}.

In \cite{Hebenstreit2010} it was shown that in the case of linearly polarized electric fields the Wigner method is equivalent to the QKT \cite{Smolyansky1997,Schmidt1998,Kohlfurst2014}, thus in that case the above equation of motion will result in the exact same spectra.

\subsection{Additional Observables}

In addition to the full one-particle distribution function, the Wigner function gives access to information about the spinor degrees of freedom of the Dirac field.
In general, the information about spin and chirality of the produced pairs can be extracted from the Wigner function.
For this we apply the corresponding projection matrices to the Wigner function and define, in analogy to \Eqref{eqn:general_f}, the projected one-particle distribution function
\begin{align}
  \label{eqn:general_f_projected}
\begin{split}
  &f_P[P\left(\mathcal{W} - \mathcal{W}_\mathrm{vac.}\right)] \\
&\qquad := \frac{1}{2\omega}\tr\left[
      P
      \left( \mathcal{W} - \mathcal{W}_\mathrm{vac.} \right)
      \left( m\mathbbm{1}+\vec{p}\cdot\vec{\gamma} \right)
      \right]\,.
\end{split}
\end{align}
By inserting \Eqs\eqref{eqn:wigner_fierz}, \eqref{eqn:wigner_vac} and \eqref{eqn:wigner_subst} into \Eqref{eqn:general_f_projected}, formulas can be derived to recover this information from the numerical results.

Let us first consider chirality.
The chiral projections are given by
\begin{align*}
  P_\mathrm{r/l} &= \frac12\left( \mathbbm{1} \pm \gamma_5 \right)\,.
\end{align*}
If the above prescription is applied, the result is
\begin{align}
  \label{eq:wigner_chiral_f}
  f_\mathrm{r/l}\definedby f[P_\mathrm{r/l}\left( \mathcal{W} - \mathcal{W}_\mathrm{vac.} \right)] \\
  &= \frac{1}{4\omega}\left( m\left( \bbs -\bbs_\mathrm{vac.}  \right)+\vec{p}\cdot\left( \vec{\bbv} - \vec{\bbv}_\mathrm{vac.} \right)\mp\vec{p}\cdot\vec{\bba} \right) \\
                               &= \frac{1}{2}  \left( f \mp \frac{1}{2\omega} \vec{p}\cdot\vec{a}  \right)\,.
\end{align}
Thus a chiral asymmetry $\delta f_\mathrm{c}$ can be defined as
\begin{align*}
  \delta f_\mathrm{c} \definedby  f_\mathrm{l}-f_\mathrm{r} =  \frac{1}{2\omega} \vec{p}\cdot\vec{a}\,.
\end{align*}
If the same approach is used for the charge or spin projections
\begin{align*}
  P_{\mp Q} &= \frac12\left( \mathbbm{1} \pm \gamma^0 \right) \\
  P_{(a,b,c)} &= \frac12(\mathbbm{1}+a\,\ii\gamma^2\gamma^3 + b\,\ii\gamma^3\gamma^1 + c\,\ii\gamma^1\gamma^2)
\end{align*}
respectively, the asymmetry has vanishing real part.
However, if both are combined to find the magnetic moment
\begin{align*}
  P_{\mu_z^\pm} &= P_Q P_{(0,0,\pm1)}+P_{-Q} P_{(0,0,\mp1)}\,,
\end{align*}
a real asymmetry $\delta f_\mathrm{\mu_z}$ can be defined as
\begin{align}
  \label{eq:wigner_mag_f}
  \delta f_\mathrm{\mu_z} \definedby  f_{\mu_z^+} - f_{\mu_z^-} = \frac{1}{2\omega}\left( m a_z+(\vec{p}\times\vec{t}\,)_z \right)\,.
\end{align}
It will be shown in section \ref{sec:interpretation} that the two spin states, as distinguished by the semiclassical method, can be reconstructed from these two asymmetries of the Wigner function.

%-------------------------------------------------------------------------------
\section{The semiclassical scattering method}\label{sec:semiclass}
%-------------------------------------------------------------------------------
The semiclassical method based on a scattering ansatz \cite{Brezin1970,Popov1971,Popov1972,Popov1973,Marinov1977,Popov2001,Kleinert2008,Kleinert2013} has recently been generalized to two-component fields with the help of the squared Dirac equation \cite{Strobel2015}. Here we reproduce the results without squaring the Dirac equation.

Note that the scattering ansatz presented in the following is exact until the approximation in \Eqref{eq:APPROX} is performed. Indeed, it is possible to construct a Riccati equation for the reflection coefficient and to solve it numerically as was done for one-component fields in \cite{Dunne2006}.
 \subsection{Decomposition of the spinor operator}
 We start from the Dirac equation
 \begin{align*}
  \left(\left[\ii  \partial_\mu-e A_\mu(x)\right]\gamma^\mu-m\right)\Psi(\vec{x},t)=0
\end{align*}
 and decompose the spinor operator as
 \begin{align*}
 \begin{split}
 \hat{\Psi}(x)=\int\frac{d\cm^3}{(2\pi)^3}\e^{\ii\vec{\cm}\vec{x}}\sum_{s=\pm1}&\left(\psi_{\vec{\cm},s}(t)\hat{a}_{\vec{\cm},s}
 +\tilde{\psi}_{\vec{\cm},s}(t)\hat{b}^\dagger_{-\vec{\cm},s}\right),% \label{eq:spinordecomp}
 \end{split}
\end{align*}
where 
\begin{align*}
\tilde{\psi}_{\vec{\cm},s}(t):=\mathcal{C}\psi_{\vec{\cm},s}(t)^*
\end{align*}
and the charge conjugation operator is given by
\begin{align*}
 \mathcal{C}=\ii\gamma^2\gamma^0.
\end{align*}
The decomposition %\eqref{eq:spinordecomp} 
follows the canonical equal-time anti-commutation relations 
\begin{align*}
 \left\{\hat{\Psi}(\vec{x},t),\hat{\pi}(\vec{y},t)\right\}=\mathbbm{1}_4\,\ii\, \delta^3(\vec{x}-\vec{y})\,,
\end{align*}
for typical Heisenberg operators following the usual anti-commutation relations
\begin{align*}
 &\left\{\hat{a}_{\vec{\cm},s},\hat{a}_{\vec{k},r}^\dagger\right\}=(2\pi)^3 \delta^3(\vec{k}-\vec{\cm})\delta_{rs}\,,\\
 & \left\{\hat{b}_{\vec{\cm},s},\hat{b}_{\vec{k},r}^\dagger\right\}=(2\pi)^3 \delta^3(\vec{k}-\vec{\cm})\delta_{rs}\,,\\
 &\left\{\hat{a}_{\vec{\cm},s},\hat{b}_{\vec{k},r}^\dagger\right\}=0,
\end{align*}
if the Wronskian condition
\begin{align}
 \sum_{s=\pm1}\left(\psi_{\vec{\cm},s}(t)\psi_{\vec{\cm},s}(t)^\dagger+\tilde\psi_{\vec{\cm},s}(t)\tilde\psi_{\vec{\cm},s}(t)^\dagger\right)=\mathbbm{1}_4 \label{eq:flatWronskian}
\end{align}
holds.
\subsection{Equations for the Bogoliubov coefficients}
For convenience we choose to work in the Weyl representation, i.e.
\begin{align*}
 \gamma^{j}=\begin{pmatrix}
             0 &\sigma^j\\
             -\sigma^j & 0
            \end{pmatrix},
&&
  \gamma^{0}=\begin{pmatrix}
             0 &I_2\\
             I_2 & 0
            \end{pmatrix}
\end{align*}
where \(\sigma^j\) are the Pauli matrices.

For two-component fields solely depending on time (\(A_\mu(x)=(0,A_x(t),A_y(t),0)\)) one can make the ansatz
\begin{align}
 \psi_{\vec{\cm},s}(t)=\mathcal{C}_s\begin{pmatrix}
                      \, m\, \psi_1^s(t)\\
                      s \,m \, \psi_2^s(t)\\
                      -s( \cm_z+s\epsilon_\perp)\, \psi_1^s(t)\\
                      ( \cm_z+s\epsilon_\perp)\, \psi_2^s(t)\\
                       \end{pmatrix}\label{eq:firstAnsatz}
\end{align}
for \(s=\pm1\).
This ansatz can be derived from that in \cite{Strobel2015} if one reconstructs the solution of the Dirac equation from the solution of its squared version found there.

We observe that \(\psi_{\vec{\cm},s}(t)\) and \(\psi_{\vec{\cm},-s}(t)\) are independent since
\begin{align*}
 \psi_{\vec{\cm},s}(t)^\dagger\cdot \psi_{\vec{\cm},-s}(t)=0\,.
\end{align*}
The solutions we will find below for \(s=\pm1\) thus represent two independent solutions to the Dirac equation.

Putting this into the Dirac equation leads to
\begin{align*}
 \i\, \dot{\psi}_1^s(t)+s\, \epsilon_\perp\psi_1^s(t)-s\, p_{x-y}(t)\psi_2^s(t)&=0\,,\\
 \i\, \dot{\psi}_2^s(t)+s\, \epsilon_\perp\psi_1^s(t)-s\, p_{x+y}(t)\psi_2^s(t)&=0
\end{align*}
where we have defined
\begin{align*}
p_{x\pm y}(t)\definedby p_x(t)\pm\i p_y(t)\,, \\
\epsilon_\perp^2\definedby \cm_z^2+m^2\,.
\end{align*}
The Wronskian condition in \Eqref{eq:flatWronskian} holds if
\begin{align}
\left|\psi_1^s(t)\right|^2+\left|\psi_2^s(t)\right|^2=1\,, \label{eq:PsiNorm}
\end{align}
and
\begin{align*}
 \mathcal{C}_s=\frac{1}{\sqrt{2\epsilon_\perp(q_z+\epsilon_\perp)}}\,.
\end{align*}
If we now set
\begin{widetext}
\begin{align}
 \psi^{s}_{1}(t)&=\frac{\sqrt{cp_{x-y}(t)}}{\sqrt{2\omega(t)}}\sqrt{cp_\parallel(t)}\left(\alpha_s(t)\frac{\e^{-\frac{\i}{2}K_s(t)}}{\sqrt{\omega(t)+s\epsilon_\perp}} +\i \beta_s(t)\frac{\e^{\frac{\i}{2}K_s(t)}}{\sqrt{\omega(t)-s\epsilon_\perp}}\right),\label{eq:ansatz1} \\
 \psi^{s}_{2}(t)&=s\frac{\sqrt{cp_{x+y}(t)}}{\sqrt{2\omega(t)}}\sqrt{cp_\parallel(t)}\left(\alpha_s(t)\frac{\e^{-\frac{\i}{2}K_s(t)}}{\sqrt{\omega(t)-s\epsilon_\perp}} -\i \beta_s(t)\frac{\e^{\frac{\i}{2}K_s(t)}}{\sqrt{\omega(t)+s\epsilon_\perp}}\right) \label{eq:ansatz2}
\end{align}
\end{widetext}
with the integrals
\begin{align}
K_s(t)\definedby K_0(t)-s K_{xy}(t)\,,\label{eq:K_s}\\
K_0(t)\definedby 2\int_{-\infty}^{t} \omega(t') dt'\,, \label{eq:Kint}\\
 K_{xy}(t)\definedby \epsilon_\perp\int_{-\infty}^{t} \frac{\dot{p}_x(t')p_y(t')-\dot{p}_y(t')p_x(t')}{\omega(t')p_\parallel(t')^2}dt' \label{eq:K_xy}
\end{align}
 and 
\begin{align*}
 p_\parallel(t)^2\definedby p_x(t)^2+p_y(t)^2
\end{align*}
we find 
\begin{align}
 \dot{\alpha}_s(t)&=\frac{\dot{\omega}_{\vec{\cm}}(t)}{2 \omega(t)}G^s_+(t) \e^{\i K_{s}(t)}\beta_s(t)\,,\label{eq:2CompAlpha}\\
 \dot{\beta}_s(t)&=\frac{\dot{\omega}_{\vec{\cm}}(t)}{2 \omega(t)}G^s_-(t) \e^{-\i K_{s}(t)}\alpha_s(t)\label{eq:2CompBeta}
\end{align}
where
\begin{align*}
 G^s_\pm(t)=\i s\frac{\epsilon_\perp}{ p_\parallel(t)}\pm \frac{\dot{p}_x(t)p_y(t)-\dot{p}_y(t)p_x(t)}{\dot{p}_x(t)p_x(t)+\dot{p}_y(t)p_y(t)}\frac{\omega(t)}{ p_\parallel(t)}\,.
\end{align*}
Using \Eqs\eqref{eq:ansatz1}--\eqref{eq:ansatz2} in  the normalization condition \Eqref{eq:PsiNorm} we find
\begin{align*}
 \left|\alpha_s(t)\right|^2+\left|\beta_s(t)\right|^2=1\,.
\end{align*}

%%%%%%%%%%%%%%%%%%%%%%%%%%%%%%%%%%%%%%%%%%%%%%%%%%%%%%%%%%%%%%%%%%%%%%%%%%%%%%%
\subsection{Momentum spectrum of produced pairs} \label{sec:SCMomSpec}
%%%%%%%%%%%%%%%%%%%%%%%%%%%%%%%%%%%%%%%%%%%%%%%%%%%%%%%%%%%%%%%%%%%%%%%%%%%%%%%
The transmission probability 
\begin{align}
 W^s(\vec{\cm}):=\lim_{t\rightarrow\infty} \left|\beta_s(t)\right|^2 \label{eq:trans}
\end{align}
can be interpreted as the number of produced electron positron pairs as a function of the momentum \(\vec{\cm}\).
Using appropriate boundary conditions \cite{Dumlu2011}
\begin{align*}
 \beta_s(-\infty)=0, && \alpha_s(-\infty)=1\,, 
\end{align*}
one can find a multiple-integral description for \(\dot{\beta}^\pm(t)\) by iteratively using \Eqs\eqref{eq:2CompAlpha} and \eqref{eq:2CompBeta} following the ideas introduced in \cite{Berry1982}. We now use the fact that the integrals are dominated by regions around the classical turning points 
\begin{align}
  \omega(t_p^{\pm}):=0 \label{eq:turningpoints}\,.
\end{align}
According to \Eqref{eq:scE} the \(t_p^{\pm}\) are found in complex conjugate pairs.
By deforming the contour we extract the singularities for the turning points for which
\begin{align}
 \Im[K_0(t_p)]<0\,. \label{eq:constraint}
\end{align}
If in the following \(t_p\) is used without the superscript \(\pm\)  it will always refer to the turning point of the pair \(t_p^\pm\) which fulfills \Eqref{eq:constraint}.
Assuming that the turning points represent singularities of order \(\nu_{t_p}\), one finds \cite{Berry1982,Strobel2015}
\begin{align}
\frac{\dot{\omega}_{\vec{\cm}}(t)}{\omega(t)}&\approx\frac{dK_0(t)}{dt}\frac{\nu_{t_p}}{\nu_{t_p}+2}\frac{1}{K_0(t)-K_0(t_p)} \label{eq:APPROX}\,.
\end{align}
One can now approximate the pre-exponential factor in each integrand in the multiple-integral series by its behavior around the poles  \(t_p\) given by \Eqref{eq:APPROX} to find 
\begin{align}
\beta_s(\infty)\approx-2\sum_{t_p}\e^{-\i K_s(t_p)} \sin\left(\frac{\pi\nu_{t_p}}{2(\nu_{t_p}+2)}\right).\label{eq:approx2}
\end{align}
This approximation is semiclassical in the sense that the exponential factor, which is not approximated, presents the leading semiclassical order.
The approximation in \Eqref{eq:approx2} breaks down if the turning points get to close to each other in the complex plane as we will detail in section \ref{sec:compare_TY}.  

Since the examples covered in the present work have simple turning points, i.e.~\(\nu_{t_p}=1\), the semiclassical momentum spectrum of \Eqref{eq:trans} takes the form
\begin{align}
 W^s_{SC}(\vec{\cm})=\left|\sum_{t_p}\e^{-\i K_s(t_p)}\right|^2 \label{eq:MomentumSpectrum}\,.
\end{align}

%-------------------------------------------------------------------------------
\section{The locally constant rotating field approximation}\label{sec:semiclass_approx}
%-------------------------------------------------------------------------------
\label{sec:Approx}
It is possible to approximate the momentum spectrum of pulsed rotating fields using the result for the rotating rectangular pulse field
\begin{align*}
 \vec{E}=E_0 \Rect\left(\frac{t}{\tau}\right)\begin{pmatrix}
                    \cos(\Omega t) \\
                    \sin(\Omega t) \\
                    0
                  \end{pmatrix},
\end{align*}
with the rectangular box function
\begin{align*}
 \Rect(x)=\Theta(x)-\Theta(x-1)\,,
\end{align*}
where \(\Theta(x)\) is the Heaviside step function.
Fields of this form can be treated analytically as shown in \cite{Strobel2015}.

The idea is to replace the field by a sum of rectangular pulses with pulse length \(\tau_0\) and different constant field strength given by the form of the pulse \(E(t)\), i.e.~replace \(E(t)\) by
\begin{align*}
 E(t)\approx \sum_{j=-\infty}^\infty E\left(\left(j+\frac12\right)\tau_0\right)\Rect
 \left(\frac{t}{\tau_0}-j\right).
\end{align*}
Now one can compute the momentum spectrum of the pair creation rate using the analytic result for the pair creation rate for each of these pulses.
The shorter the length \(\tau_0\) the better becomes this approximation.
Using that for the rectangle pulse the only turning points which contribute are those whose real part lies within the pulse range, it is possible to perform the limit \(\tau_0\rightarrow0\) which leads to
\begin{align}
 W^s_\text{approx}(\vec{\cm})=\left|\sum_{j=-\infty}^\infty \e^{K_s\left(\vec{\cm},{E\left(\Re[t_j]\right)}\right)}\right|^2 ,\label{eq:approx}
\end{align}
where \(K_s(\vec{\cm},E)\) is the integral from \Eqref{eq:K_s} which is, for the constant rotating field, given by \Eqs\eqref{eq:Krot} and \eqref{eq:K_xyrot} and \(t_j\) are the turning points given in \Eqref{eq:rotatingcompltp}.

The LCRFA approximates the field by a constant rotating field at every time. 
Therefore effects from the time variation caused by the shape of the pulse are neglected with respect to the effects of the rotation.
Accordingly the approximation is reasonable for long enough pulses in which the timescale of the rotation \(1/\Omega\) is smaller than the time scale of the pulse \(\tau\), i.e.~\(\sigma:=\Omega\tau\gg1\).

%-------------------------------------------------------------------------------
\section{Comparison of the Methods}\label{sec:cmp}
%-------------------------------------------------------------------------------
\begin{figure}
 \centering
 \includegraphics{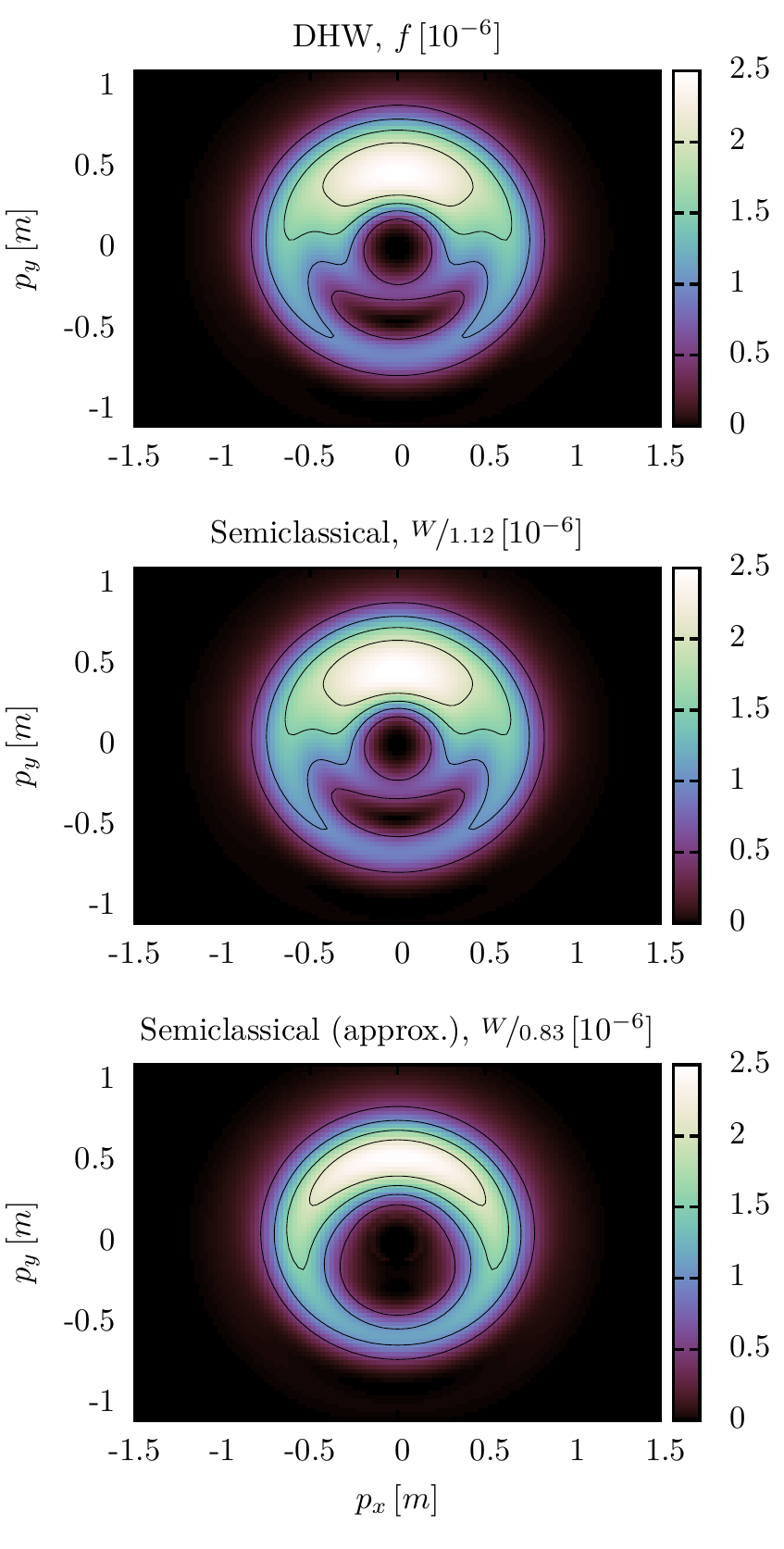}
 \caption{\label{fig:spectra_methods}
  Momentum spectrum of the Sauter pulse for \(\tau=10/m\), \(\sigma=6\) and \(\epsilon=0.1\).
  Top: The Wigner result.
  Middle: The semiclassical result divided by 1.12.
  Bottom: The result of the LCRFA divided by 0.83.
}
\end{figure}
In this section we compare the three methods we described above. 
We do so for the example of the rotating pulse in \Eqref{eqn:puls-sauter-rot} where we choose the pulse to have the shape of a Sauter pulse
\begin{align} 
 E(t)=\frac{\epsilon E_\text{c}}{\cosh^2\left(\nicefrac{t}{\tau}\right)}\,,&&E_\text{c}:=\frac{m^2}{e}\,, \label{eq:ESauter}
\end{align}
where we defined the electric field in units of the critical electric field
\begin{align*}
 \epsilon=\frac{E_0}{E_\text{c}}\,.
\end{align*}
This pulse has been studied before with the Wigner method in \cite{Blinne2013}.
It has the advantage that the limit to the non-rotating pulse \(\Omega\rightarrow0\) can be treated analytically with both the Wigner method and the scattering approach (see appendix \ref{app:Sauter} for more details).

As for the constant rotating field discussed in \cite{Strobel2015} we find that there is an infinite amount of turning points for the rotating Sauter pulse. But in contrast to this case the turning points in the general case have different real as well as different imaginary parts (see \Figref{fig:tp_numeric} for a plot of the turning points). 
This would in principle require a separate treatment of all of them. 
However the closer a pair of turning points is to the real axis, the bigger is its influence on the pair creation rate \cite{Dumlu2011}, such that it is sufficient to study a finite number of turning points to have a good approximation for the pair creation rate (see appendix \ref{sec:numsemi} for details). 
Note that this holds true also within the LCRFA where it is sufficient to evaluate the sum in \Eqref{eq:approx} up to a finite \(|j|\).

Proceeding numerically gives us the possibility to compare the momentum spectrum calculated with the help of the Wigner method and the numerical and LCRFA semiclassical results in section \ref{sec:compare_MS}.
For a more quantitative comparison we compute the total pair creation rate and compare the three methods concerning the result and the computation time in section \ref{sec:compare_TY}.
An interpretation of the two solutions found for the scattering method in the light of the Wigner method is presented in section \ref{sec:interpretation}

%-------------------------------------------------------------------------------
\subsection{Comparison of momentum spectra}\label{sec:compare_MS}
%-------------------------------------------------------------------------------
\begin{figure}[tpb]
 \centering
 \includegraphics{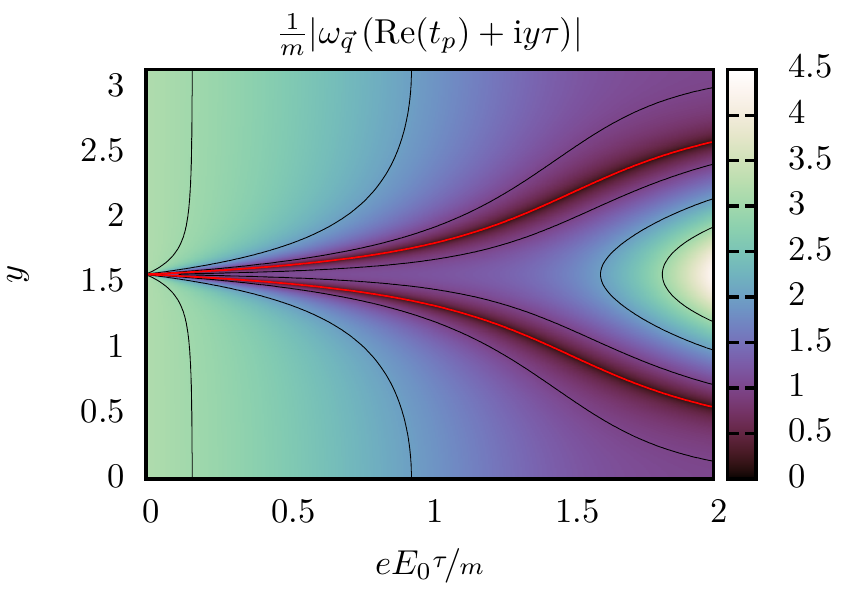}
 \caption[Value of \(|\omega(t)|\) for \(t=s_j+\ii y \tau\) depending on  \(e E_0 \tau/m\)]{Value of \(|\omega(t)|\) for \(t=\Re[t_p]+\ii y \tau\) depending on  \(e E_0 \tau/m\) for \(\cm_x=3m,\,\cm_y=\cm_z=0\). We see that for small \(e E_0 \tau/m\) the turning points (red line) get closer and the assumption that \Eqref{eq:APPROX} holds for every turning point is not satisfied anymore.}
 \label{fig:TPSauter}
\end{figure}
In order to compare the momentum spectra calculated by all three methods, let us choose an illustrative example.
Let $\tau=10/m$ and $\sigma=6$.
The Wigner method momentum spectrum of this pulse has already been published in \cite{Blinne2013}.
\Figref{fig:spectra_methods} shows the spectrum as computed by all the available methods.
It turns out that the semiclassical method overestimates the pair production probability by roughly 12 percent as compared to the result of the Wigner method.
The result in LCRFA has the same order of magnitude as the other results but underestimates certain features of the momentum spectrum.
We will see in the next section that this is due to the small number of field cycles ($\sigma=6$) and that the approximation gets better for bigger \(\sigma\).
%-------------------------------------------------------------------------------
\subsection{Comparison of the total particle yield}\label{sec:compare_TY}
%-------------------------------------------------------------------------------
\begin{figure*}[htbp]
 \includegraphics{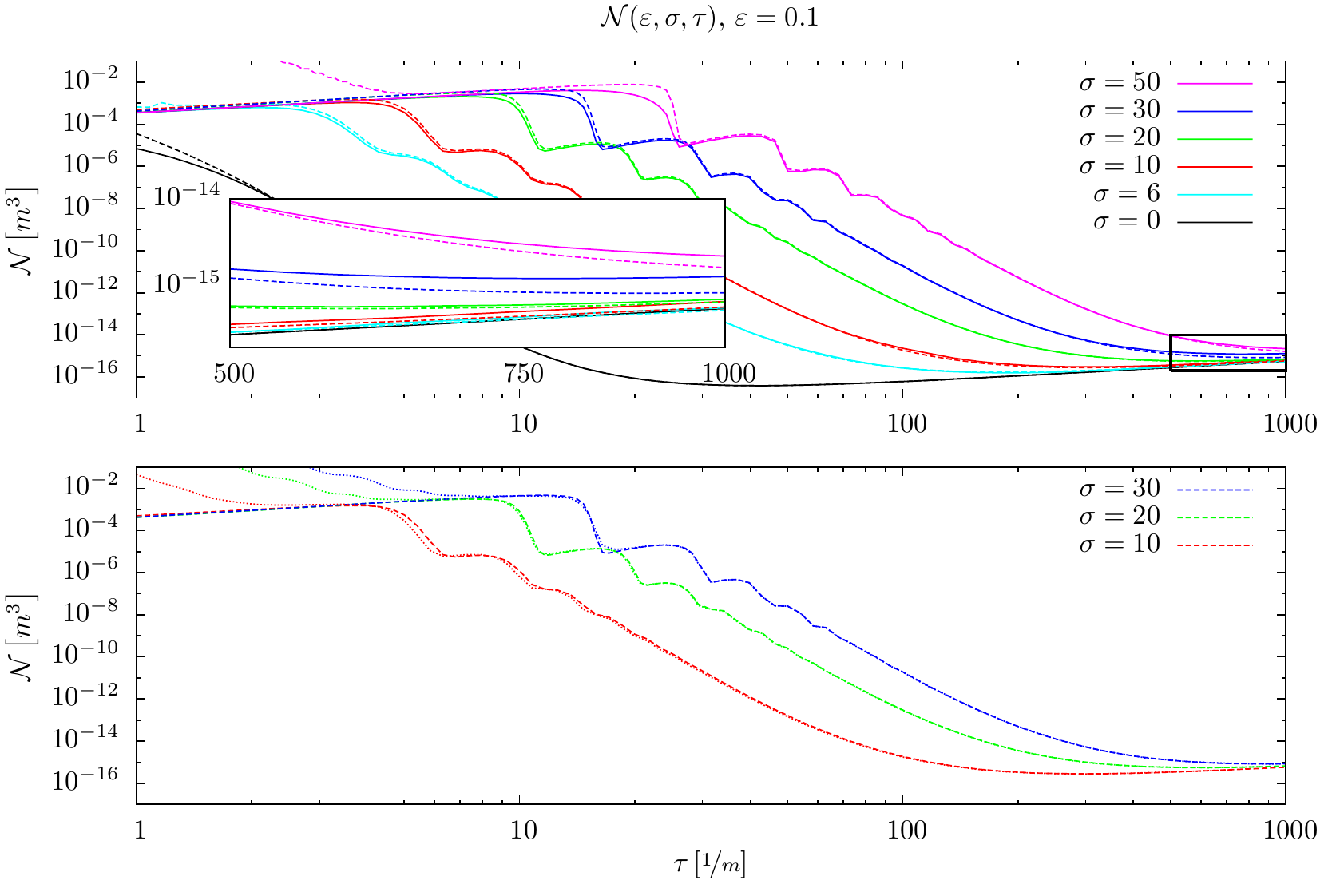}
 \caption{\label{fig:total_yield}Comparison of the total particle number per Compton volume of the rotating Sauter pulse for \(\epsilon=0.1\) as a function of the pulse length \(\tau\).
 Top:
 Solid lines show particle yield as calculated using the Wigner method, dashed lines show particle yield as calculated using semiclassical method.
 In the cases $\sigma\in\{6,10\}$ a noise suppression method has been used when integrating over the spectra of the Wigner method to obtain the 3D totals.
 For pulse durations of the order of $10$ Compton times the semiclassical method tends to overestimate the pair production rate, especially just below the resonances.
 For pulse durations approaching 1000 Compton times or more, the numerical difficulties of the Wigner method become apparent.
 Bottom:
 Dashed lines show the particle yield as calculated using the numerical semiclassical method, dotted lines show particle yield as calculated using the LCRFA.
 One finds that for long enough pulses the approximation agrees with the numerical results.
 For short pulses the approximation that the field is locally constant gets worse.
 The fixed number of turning points of 9 used for this example in difference to the adaptive algorithm used for the numerical method contributes to the error as well.%
}
\end{figure*}
\begin{figure*}[htbp]
 \includegraphics{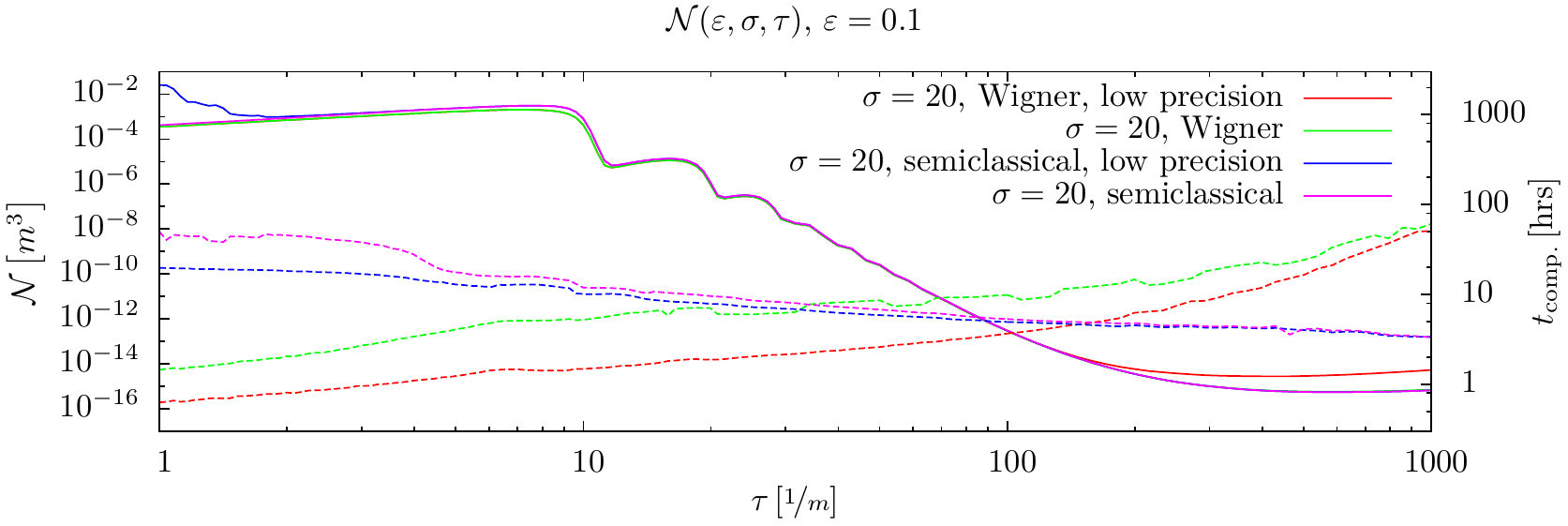}
 \caption{\label{fig:total_yield_20}Comparison of the total particle number per Compton volume of the rotating Sauter pulse for \(\epsilon=0.1,\,\sigma=20\) as a function of the pulse length \(\tau\) for different settings regarding precision.
 Solid lines show particle yield, dashed lines show processor time per spectrum.
 The Wigner method increases in computation time strongly towards longer pulse durations until it becomes numerically unfeasible.
 The semiclassical method however becomes even numerically cheaper for longer pulses.
}
\end{figure*}
We can compute the total particle yield per volume from the momentum spectrum by integrating over momentum space
\begin{align*}
 \frac{\Gamma_s}{V}:=\int  \frac{d^3\cm}{(2\pi  )^3}\, W_s(\vec{\cm})\,.
\end{align*}
Comparing the results we find that the methods agree for an intermediate range of pulse length \(\tau\) (see \Figref{fig:total_yield}).

We also find that for short pulses the semiclassical method is not working anymore. 
This can be explained by looking at the non-rotating Sauter pulse which is studied in more detail in appendix \ref{app:Sauter}. 
Taking the turning points given by \Eqref{eq:TPSauter} into consideration we find that for decreasing \(eE_0\tau/m\) the turning points get closer in the complex plane (see \Figref{fig:TPSauter} for a plot of \(|\omega(t)|\) around the turning points).
The approximation performed in section \ref{sec:semiclass} assumes that \Eqref{eq:APPROX} holds for every turning point. 
This is not the case if the different turning points get too close to each other in the complex plane.
We thus find that the semiclassical approximation breaks down for the Sauter pulse for short pulses which can be seen in the total pair creation rate in \Figref{fig:total_yield}.
There we also see that the same happens for the rotating Sauter pulse.

For longer pulses the Wigner method becomes numerically challenging.
This is due to the fact that the integration from \(t=-\infty\) to \(t=\infty\), which is performed analytically in the semiclassical method, needs more steps the longer the pulse becomes.
For too long pulses the precision of the result is limited by computational errors (see \Figref{fig:total_yield_20} for a comparison of the computation times of the numerical methods).
We find that for \(\sigma=20\) both numerical methods have a comparable computation time around \(\tau\sim40-100\,m\). 
For shorter pulses the Wigner method is computationally faster while for longer ones the semiclassical method should be preferred.

The computation time, when using the LCRFA, is dominated by the integration over the momentum spectrum which has to be performed numerically. 
We find that the approximation gets better for longer pulse length \(\tau\) and higher \(\sigma\) (see \Figref{fig:total_yield}, observe that within the LCRFA calculations the number of considered turning points has been fixed to 9 in contrast to the adaptive method used for the numerical method described in appendix \ref{sec:numsemi}).
This can be explained by the fact, that the longer the pulse, the better gets the approximation of the pulse being locally constant and that for higher \(\sigma\) the effects of the rotation become more important with respect to the effects of the pulse shape since there are more rotation cycles per pulse length.

%-------------------------------------------------------------------------------
\subsection{Interpretation of the two solutions of the semiclassical method}\label{sec:interpretation}
%-------------------------------------------------------------------------------
In section \ref{sec:semiclass} we found two independent solutions of the Dirac equation which were interpreted as different spin components in \cite{Strobel2015}.
To compare these with the Wigner method we can construct the projections
\begin{align}
 P_s \definedby \frac{1}{2}\left(\mathbbm{1}+s\gamma_5\right)-s\frac{1}{2}\gamma_5\frac{(\cm_z+\epsilon_\perp)\mathbbm{1}+\gamma^3}{\epsilon_\perp} \nonumber
 \\
 \label{eq:projector_interpret}
 &= \frac{1}{2}\mathbbm{1} +s\frac{1}{2}\left( \frac{\cm_z}{\epsilon_\perp} (P_l-P_r) + \frac{m}{\epsilon_\perp}\left( P_{\mu_z^+} - P_{\mu_z^-} \right)\right)
\end{align}
for \(s=\pm1\).
These projections are idempotent
\begin{align*}
 P_s\cdot P_s=P_s\,, 
\end{align*}
and orthogonal 
\begin{align*}
 P_s\cdot P_{-s}=0\,,
 \\
 P_s+ P_{-s}=\mathbbm{1}\,.
\end{align*}
They also fulfill 
\begin{align*}
 P_s\cdot \psi_{\vec{\cm},s}=\psi_{\vec{\cm},s}\,,&&
 P_s\cdot \psi_{\vec{\cm},-s}=0
\end{align*}
for the  two independent solutions of the Dirac equation defined in \Eqref{eq:firstAnsatz}.
Accordingly they project on the two parts of the spectrum which correspond to these solutions.
In \Eqref{eq:projector_interpret} it is evident, that the two solutions from the scattering method correspond to a linear combination of chirality and magnetic momentum.

\begin{figure*}[p]
 \includegraphics[width=\linewidth]{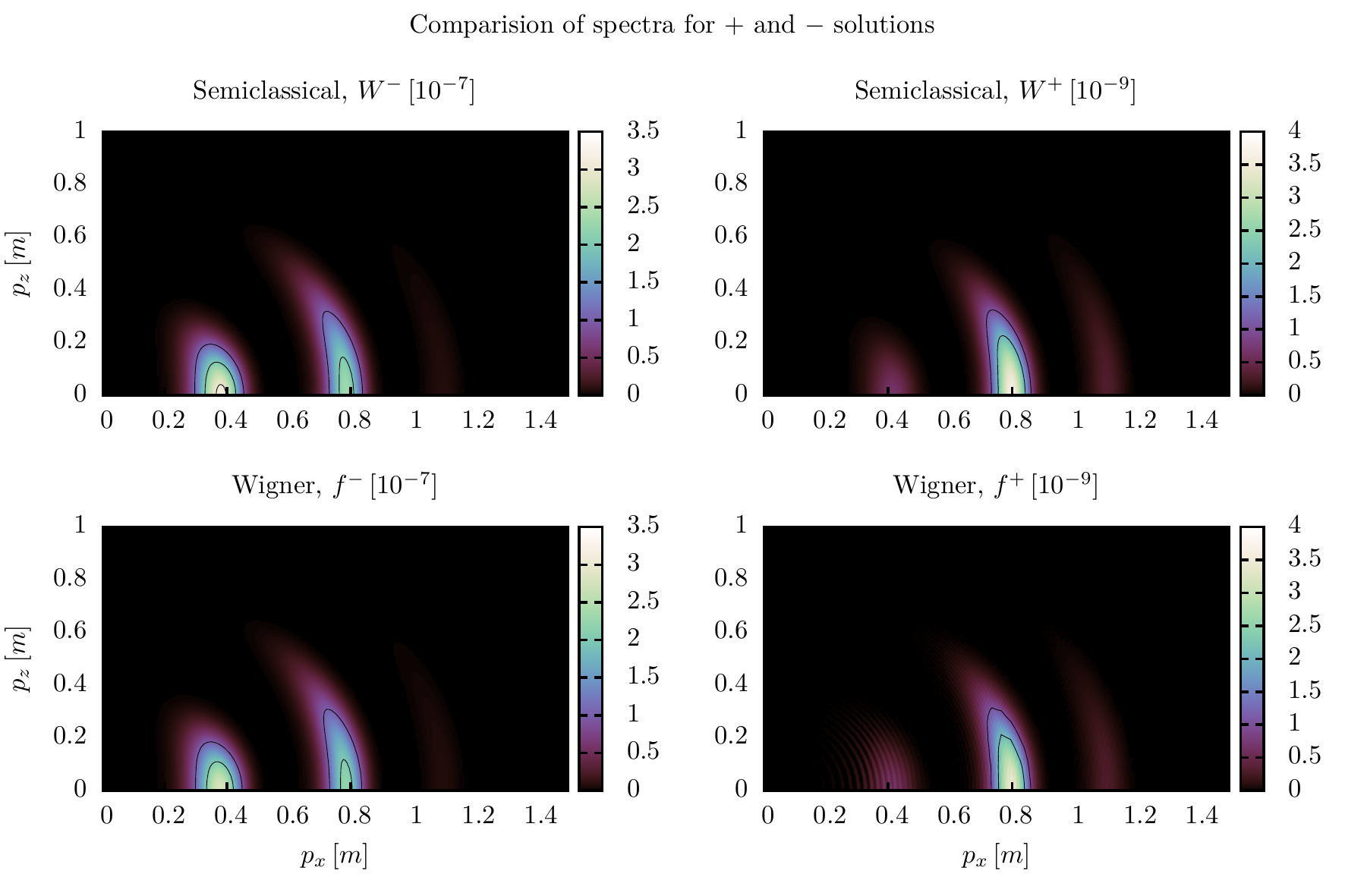}
 \caption{\label{fig:spectra_spin_cmp}Comparison of spectra for semiclassical $+$ and $-$ solutions as well as corresponding spectra from the Wigner method.
 The pulse parameters are $\tau=46.42$, $\sigma=20$. Except for an interference pattern around \(p_x\sim0.4\) in \(f^-\), which is not found in the semiclassical results, the spectra of the two methods agree with each other.}
\end{figure*}

\begin{figure*}[p]
 \includegraphics{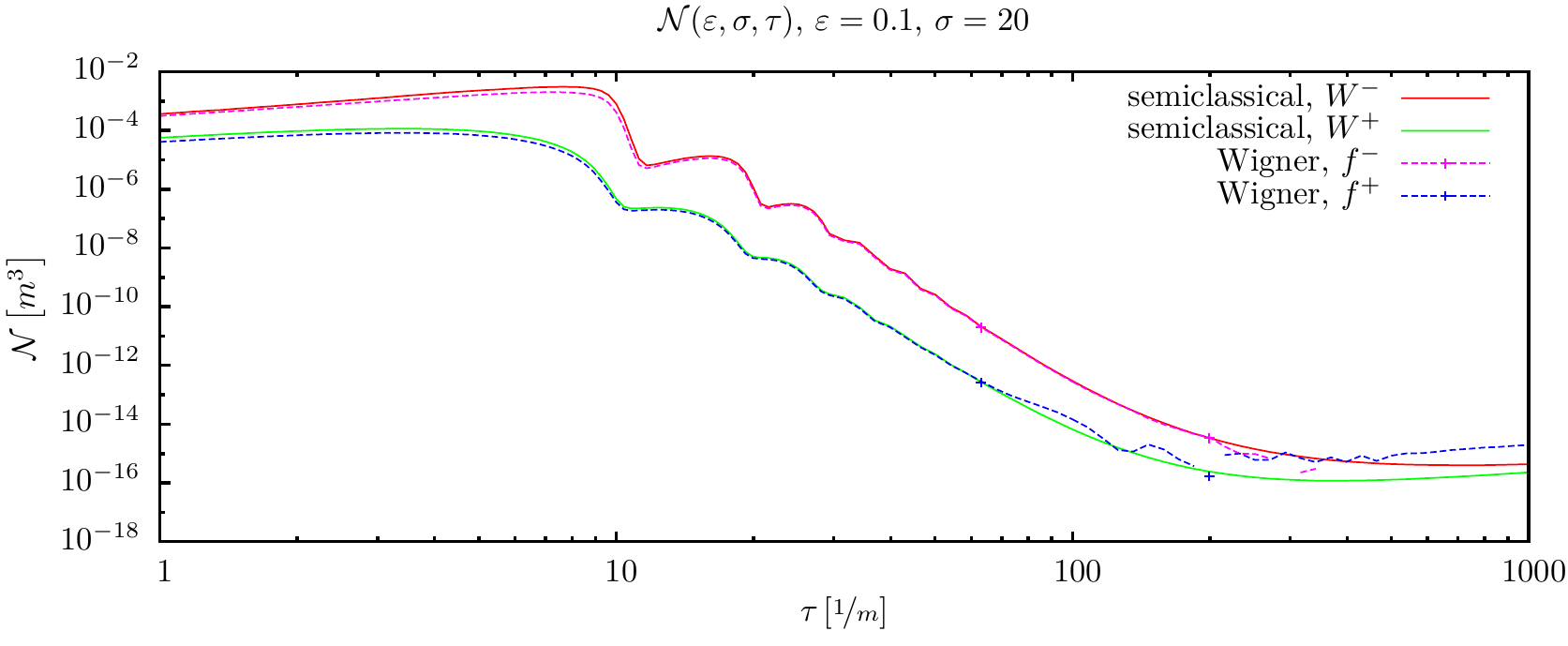}
 \caption{\label{fig:spin_plot_20}Comparison of the particle yield for the semiclassical $W^+$ and $W^-$ solutions with the corresponding Wigner function projections $f^+$ and $f^-$.
 The Wigner method data suffer from a lack of precision for higher \(\tau\).
 The `+' data points have been calculated using higher precision.
 We find that the results agree with each other.
}
\end{figure*}

While in the context of the Wigner function, which contains the full spinor information, \(f_\mathrm{c}\) and \(f_{\mu_z}\), given in \Eqs\eqref{eq:wigner_chiral_f} and \eqref{eq:wigner_mag_f} respectively, are the physically more relevant observables, we will construct \(f^s\) to show the connection to the solutions of the semiclassical method.
The projected one-particle function, as defined in \Eqref{eqn:general_f_projected}, using this projection is given by
\begin{align*}
 f^s&= f\left[P_s\left( \mathcal{W} - \mathcal{W}_\mathrm{vac.} \right)\right]\\
&= \frac{1}{2} \left( f+s \, \delta f_\mathrm{sc} \right)
\end{align*}
with the corresponding asymmetry $\delta f_\mathrm{sc}$.
The latter can be related to the previously defined chiral and magnetic momentum asymmetries $\delta f_\mathrm{c}$ and $\delta f_{\mu_z}$ respectively as
\begin{align*}
 \delta f_\mathrm{sc} &= \frac{q_z}{\epsilon_\perp}\delta f_\mathrm{c}+\frac{m}{\epsilon_\perp}\delta f_{\mu_z}\,.
\end{align*}
Using this we find that the data for the semiclassical and Wigner method agree (see Figs.~\ref{fig:spectra_spin_cmp} and \ref{fig:spin_plot_20}). 
This shows that the two independent solutions of the semiclassical method represent spinor states with specific chirality and magnetic moment. 

%-------------------------------------------------------------------------------
\section{Conclusion}\label{sec:concl}
%-------------------------------------------------------------------------------
In this work we compare the semiclassical pair creation rate found using a scattering ansatz to the rate resulting from the Wigner method.
The numerical semiclassical results are found to be complementary to the Wigner results in terms of accuracy as well as computation time.
While the results agree for intermediate pulse lengths, for short pulses the semiclassical approximation breaks down and the computation time becomes high.
The advantage of the semiclassical method is that one does not have to integrate numerically with respect to real time.
This is especially useful for long pulse lengths since numerical problems arise in the Wigner method, which increase the required computation time  and limit accuracy.

We also introduce the LCRFA for rotating field pulses and show that is has the same features as the numerical results and works especially well for pulse lengths \(\tau\gg\nicefrac{1}{m}\).
This is intriguing since the parameters of current and near future laser systems fulfill this requirement.
It can therefore be  used to study different pulse shapes qualitatively. This is of special interest since it was shown that pair creation depends sensitively on the pulse shape of lasers \cite{Schutzhold2008,Dunne2009,Bell2008,DiPiazza2009,Monin2010,Monin2010B,Heinzl2010,Bulanov2010}.
Using an optimization approach similar to the one used in \cite{Hebenstreit2014} one could use the LCRFA to investigate a great number of possible pulse shapes because of the small computation time compared to numerical methods.
The latter could be used subsequently to verify the results for interesting pulse shapes.

In addition to that we find that the two independent solutions of the Dirac equation found for the scattering approach, which have been interpreted as spin states in \cite{Strobel2015} can be given a clear interpretation. 
This is done with the help of projections which show that the solutions can be associated with particles having a specific chirality and magnetic moment and how the same spectra can be computed within the Wigner method.
As mentioned in \cite{Strobel2015} the fact that one of these solutions dominates the spectrum for a certain range of parameters might help to differentiate pairs produced by the Schwinger process from other particles detected in high energy experiments.

\acknowledgments

The authors thank Holger Gies and She-Sheng Xue for many fruitful discussions. 
ES acknowledges discussions with Carlos Arg\"uelles and Clément Stahl.  
ES is supported by the Erasmus Mundus Joint Doctorate Program by Grant Number 2012-1710 from the EACEA of the European Commission.
AB acknowledges discussions with J. Borchardt, C. Kohlfürst, S. Lippoldt, D. Schinkel and N. Seegert.
AB is supported by the DFG under grants GRK 1523/2, and SFB-TR18.

\appendix

%-------------------------------------------------------------------------------
\section{Numerical aspects of the Wigner method}\label{sec:numwigner}
%-------------------------------------------------------------------------------
The Runge--Kutta--Cash--Karp--54 implementation \cite{Ahnert2011} as used in this work features an automatic step size control.
The implementation accepts two parameters, $\mathrm{abserr}$ and $\mathrm{relerr}$ and then chooses step size such that it ensures the approximate integration error to be smaller than
\[
\varepsilon = \mathrm{abserr} + |\mathbf{x}|\,\mathrm{relerr}\,.
\]
Previous experience \cite{Blinne2013} showed, that $\mathrm{relerr}$ should be set to 0, because of big intermediate function values, which would spoil the overall precision.
As a result the only external parameter to the numerical calculations is the absolute error tolerance $\mathrm{abserr}=10^{-k}$.
When using double precision with the fast-math compiler option values for $k$ up to 14 can be used.

%-------------------------------------------------------------------------------
\section{Numerical aspects of the semiclassical method}\label{sec:numsemi}
\begin{figure*}[htb]
 \includegraphics{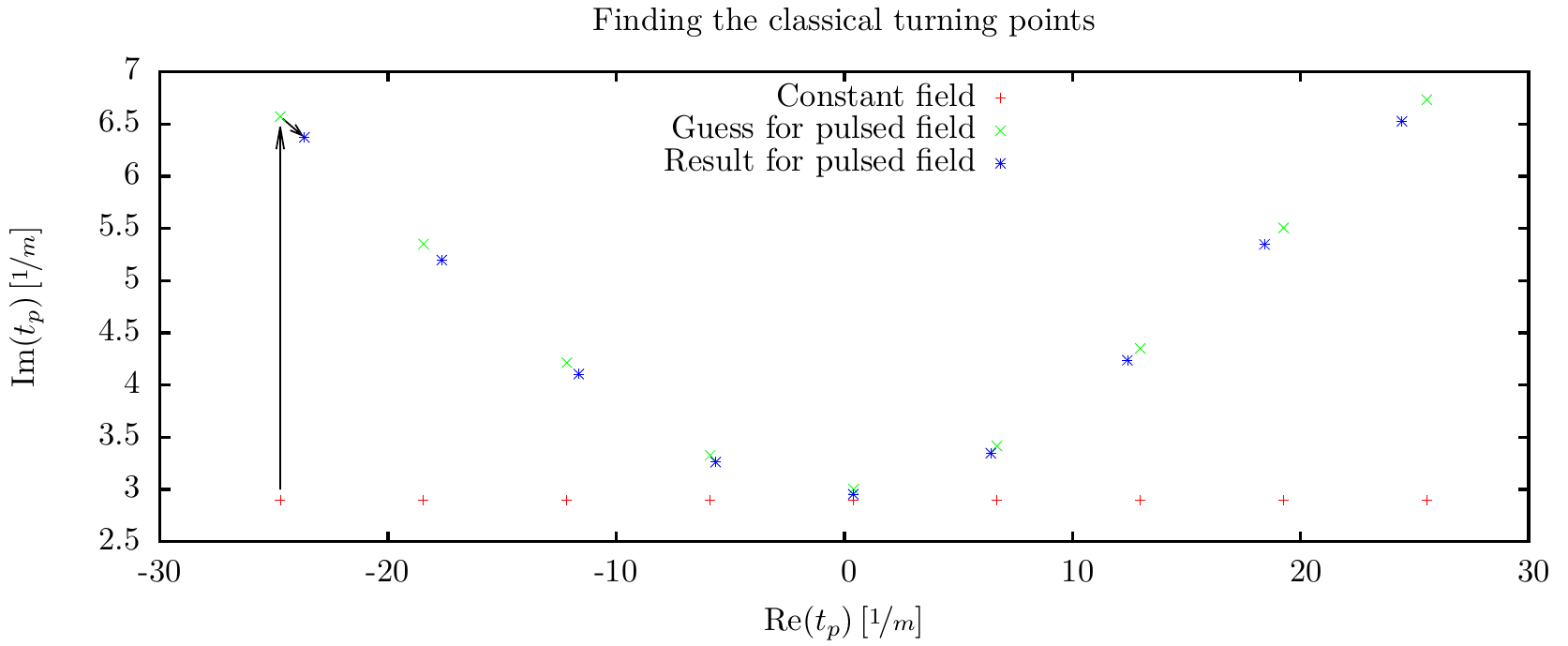}
 \caption{\label{fig:tp_numeric}%
This figure shows how the turning points for the pulsed rotating field are found.
The turning points of the constant rotating field are shifted away from the real axis by replacing the field strength parameter by the pulse shape (upward arrow).
Afterwards the correct turning points can be found by a numerical search (smaller arrow).%
}
\end{figure*}

%-------------------------------------------------------------------------------
In order to calculate the semiclassical pair production rates, first the classical turning points of the given potential need to be found.
This is done by numerically solving $\omega(t_p)=0$ for complex $t_p$ using a Newton-Raphson method, which needs an initial guess that is in some sense close to the desired solution.
The known turning points for the constant rotating field discussed in appendix  \ref{app:constrot} and given in \Eqref{eq:rotatingcompltp} may be used as a starting point.
Unfortunately, these points are too far away from the desired solutions to have them found reliably by the numerical search.
If however the field strength parameter $E_0$ in \Eqref{eq:rotatingcompltp} is replaced by the pulse shape $E(t_j)$ 
\begin{align}
 E_0\rightarrow E(t_j)=E_0 \frac{1}{\cosh(\Re(t_j))^2}\,, \label{eq:replacement}
\end{align}
the result is a good enough starting point 
(see \Figref{fig:tp_numeric} for a depiction of this behavior).
In fact these are the turning points of the LCRFA which is studied in section  \ref{sec:semiclass_approx}.
In this way we also get a nomenclature for the turning points, by giving them the same name as the corresponding ones of the constant rotating field.

For the computation the momentum grid is divided into several parts for parallelization.
For each of these parts the number of used pairs of turning points is chosen adaptively.
Starting from \(t_0\) (which is the pair closest to \(t_0\) given by using \Eqref{eq:replacement} in \Eqref{eq:rotatingcompltp}) \(t_j\) and \(t_{-j}\) are computed until the contribution of the pairs to \(W_s(\vec{\cm})\) is less than \(0.1\%\).

The semiclassical method heavily relies on integrals in the complex plane.
These are being reworded into multiple real integrals by parameterization of the integration paths.
Afterwards the GNU Scientific Library is used to carry out the numerical integrals, specifically adaptive Gauss--Kronrod and Clenshaw--Curtis rules are being used.
The adaptive algorithms are also tuned by an absolute and a relative error tolerance.
Still, it is necessary to evaluate the vector potential for complex times.
The indefinite integral of the field given by \Eqs\eqref{eqn:puls-sauter-rot} and \eqref{eq:ESauter} can be given as
\begin{widetext}
\begin{align}
  \label{eqn:field-integral}
  \int \vec{E}(t)\,dt
   &= \frac{E_0\tau}{2} \cdot
   \begin{pmatrix}
      e^{-it\Omega} H_1+e^{it\Omega}H_2-e^{\frac{t}{\tau}(2-i\tau\Omega)}\frac{\tau\Omega}{2i+\tau\Omega}H_3+e^{\frac{t}{\tau}(2+i\tau\Omega)}\frac{\tau\Omega}{2i-\tau\Omega}H_4
      +2\cos(t\Omega)\tanh(\frac{t}{\tau})
      \\
      ie^{-it\Omega} H_1-ie^{it\Omega}H_2+e^{\frac{t}{\tau}(2-i\tau\Omega)}\frac{\tau\Omega}{-2+i\tau\Omega}H_3-e^{\frac{t}{\tau}(2+i\tau\Omega)}\frac{\tau\Omega}{2+\tau\Omega}H_4
      +2\sin(t\Omega)\tanh(\frac{t}{\tau})
      \\
      0
   \end{pmatrix}
\end{align}
\end{widetext}
with
\[
\begin{array}{rrrl}
 H_1 = \prescript{}{2}{F}_1\big( 1,&   -\frac{i}2 \tau\Omega,& 1-\frac{i}{2}\tau\Omega,& -e^{\frac{2t}{\tau}} \big) \\
 H_2 = \prescript{}{2}{F}_1\big( 1,&   \frac{i}2 \tau\Omega,& 1+\frac{i}{2}\tau\Omega,& -e^{\frac{2t}{\tau}} \big) \\
 H_3 = \prescript{}{2}{F}_1\big( 1,& 1-\frac{i}2 \tau\Omega,& 2-\frac{i}{2}\tau\Omega,& -e^{\frac{2t}{\tau}} \big) \\
 H_4 = \prescript{}{2}{F}_1\big( 1,& 1+\frac{i}2 \tau\Omega,& 2+\frac{i}{2}\tau\Omega,& -e^{\frac{2t}{\tau}} \big)\,,
\end{array}
\]
where $\prescript{}{2}{F}_1$ denotes the Gaussian hypergeometric function.
Due to the singularities at solutions of $\cosh^2\left(\nicefrac{t}{\tau}\right)=0$, the vector potential must have branch cut discontinuities.
These singularities are found on the imaginary axis at
\[
t_\mathrm{k} = \frac{2k+1}{2}\pi\,\tau\,.
\]
The form given in \Eqref{eqn:field-integral} is discontinuous on straight lines that start at the singularities and continue parallel to the real axis towards positive real infinity.
However in the region with negative real and positive imaginary part the potential given in \Eqref{eqn:field-integral} is continuous.
By exploiting the symmetries of the electric field $E_\mathrm{x/y}(t)\to\pm E_\mathrm{x/y}(-t)$, this continuous portion can be carefully mirrored towards the right hand side of the imaginary axis leaving all the discontinuities strictly on the imaginary axis.
Finally the evaluation of the Gaussian hypergeometric function with complex arguments is left to a code described in \cite{Michel2008} which is available under the name AEAE at the Computer Physics Communications Program Library.
\newline % fixes broken alignment between widetext-environments
%-------------------------------------------------------------------------------
\section{Results for the constant rotating pulse}\label{app:constrot}
%-------------------------------------------------------------------------------
The integrals of \Eqs\eqref{eq:Kint}--\eqref{eq:K_xy} at the turning points for the constant rotating field have been computed beforehand in \cite{Strobel2015}. Here we reproduce them for the sake of completeness. The turning points and the integrals can be found to be
\begin{widetext}
\begin{align}
 \Omega t_k^\pm=\arcsin\left(\frac{\cm_x}{\cm_\parallel}\right)\pm\i\,\operatorname{arcosh}\left(\frac{\left( \cm_\parallel^2+\epsilon_\perp^2\right)\left(\frac{\Omega}{m\epsilon}\right)^2+m^2}{2\left(\frac{\Omega}{m\epsilon}\right)\,  \cm_\parallel\, m}\right)+2\pi k\,, \label{eq:rotatingcompltp}
\end{align}
 \begin{align}
 K(\vec{\cm},E_0)%=K(t_k)
 &=\i\frac{4\epsilon_\perp}{\Omega}\sqrt{1-y_+^2}\left[\E\left(\sqrt{\frac{1-y_-^2}{1-y_+^2}}\right)-\K\left(\sqrt{\frac{1-y_-^2}{1-y_+^2}}\right)-2k\,\i \E\left(\sqrt{\frac{y_-^2-y_+^2}{1-y_+^2}}\right)\right]+\Phi\,, \label{eq:Krot}
\end{align}
\begin{align}
\begin{split}
 K_{xy}(\vec{\cm},E_0)%=K_{xy}(t_k)
 =&-\frac{\i g}{\sqrt{1-y_+^2}}
 \left[\K\left(\sqrt{\frac{1-y_-^2}{1-y_+^2}}\right)-y_-y_+\PiI\left(1-y_-^2,\sqrt{\frac{1-y_-^2}{1-y_+^2}}\right)\right]+\Phi_{xy}\\
 &+2k\frac{\,g}{\sqrt{1-y_+^2}}\left[\left(1-y_-y_+\right)\K\left(\sqrt{\frac{y_-^2-y_+^2}{1-y_+^2}}\right)+({y_-^2-1})\frac{y_+}{y_-}\PiI\left(\frac{1}{y_-^2}\frac{y_-^2-y_+^2}{1-y_+^2},\sqrt{\frac{y_-^2-y_+^2}{1-y_+^2}}\right)\right]\,, \label{eq:K_xyrot}
 \end{split}
\end{align}
\centering{where}
\begin{align*}
 y_{\pm}:=\i\frac{\Omega\,\cm_\parallel\pm e E_0 }{\Omega\,\epsilon_\perp}\,,&&
\cm_\parallel:=\sqrt{\cm_x^2+\cm_y^2}\,.
\end{align*}
\end{widetext}
The quantities \(\Phi\) and \(\Phi_{xy}\) are physically-irrelevant global phases.

%%%%%%%%%%%%%%%%%%%%%%%%%%%%%%%%%%%%%%%%%%%%%%%%%%%%%%%%%%%%%%%%%%%%%%%%%%%%%
\section{Analytic calculation of the momentum spectrum for the Sauter pulse} \label{app:Sauter}
%%%%%%%%%%%%%%%%%%%%%%%%%%%%%%%%%%%%%%%%%%%%%%%%%%%%%%%%%%%%%%%%%%%%%%%%%%%%%
In this appendix we want to calculate the integral \(K_0(t)\) for the non-rotating Sauter pulse which is given by \Eqs\eqref{eqn:puls-sauter-rot} and \eqref{eq:ESauter} for \(\Omega=0\).
We start from the potential
\begin{align*}
 A_\mu(x)&=(0,A(t),0,0),\\ 
 A(t)&=eE_0\tau \left[1+\tanh\left(\frac{t}{\tau}\right)\right].
\end{align*}
The turning points as defined in \Eqref{eq:turningpoints} can be found to be
\begin{align}
 t_j^\pm&=\tau \artanh\left[\frac{\cm_x\pm\ii\tilde{\epsilon}_\perp}{e E_0 \tau}-1\right]+\ii \pi j \tau\,, \label{eq:TPSauter}
\end{align}
for \(j\in\mathbb{N}\) with
\begin{align*}
\tilde{\epsilon}_\perp^2\definedby \epsilon_\perp^2+\cm_y^2\,.
\end{align*}
This means we find an infinite number of turning points which all have the same real part
\begin{align*}
 s_j=\Re(t_j^\pm)=\frac{1}{4}\log\left(\frac{\cm_x^2+\tilde{\epsilon}_\perp^2}{(\cm_x-2eE_0\tau)^2+\tilde{\epsilon}_\perp^2}\right).
\end{align*}
The integral from \Eqref{eq:Kint} gives
\begin{widetext}
 \begin{align}
\begin{split}
 K_0(t)=&-\tau\frac{2}{\gamma}\log\left[\frac{\gamma}{m}(\omega(t)+\cm_x)+\tanh\left(\frac{t}{\tau}\right)\right]
    -\tau\sum_{l=\pm1}l\cm_l\Bigg(\log\left[1-l\tanh\left(\frac{t}{\tau}\right)\right]
    \\&\hspace{1.5cm}
    -\log\left[\left(\frac{\gamma}{m}\cm_x+l\right)\left(\frac{\gamma}{m}\cm_x+\tanh\left(\frac{t}{\tau}\right)\right)+\frac{\gamma^2}{m^2}\left(\cm_l\omega(t)+\tilde{\epsilon}_\perp^2\right)\right]\Bigg)+\tilde{\Phi}\label{eq:Kint3}\,,
\end{split}
 \end{align}
\end{widetext}
where \(\tilde{\Phi}\) is a physically-irrelevant global phase and we introduced the Keldysh parameter for the pulse length \(\tau\) which is defined as 
\begin{align*}
 \gamma:=\frac{m }{e E_0\tau}
\end{align*}
and we also defined
\begin{align*}
 \cm_\pm\definedby \sqrt{\left(  \cm_x-\frac{ms}{\gamma}(1\pm1)\right)^2+\tilde{\epsilon}_\perp^2}\,.
\end{align*}
Using the explicit form of the turning points of \Eqref{eq:TPSauter} and assuming \(E_0>0\) we find that the imaginary part of the integral from \Eqref{eq:Kint} at the turning points is given by
\begin{align*}
\Im[K_0(t_j^\pm)]=&\mp\frac{\pi}{2} \tau \frac{1}{\gamma}\left(\gamma \cm_++\gamma \cm_--2m\right)\,.
\end{align*}
According to the condition in \Eqref{eq:constraint} only the turning points for which the imaginary part is negative contribute.
That still leaves an infinite number of turning points which will give the same contribution to the sum in \Eqref{eq:MomentumSpectrum}.
However \Eqref{eq:MomentumSpectrum} only holds if the turning points have a different real part.
This is connected to how the contour is deformed to extract the contributions of the poles.
We chose the contour such that it follows the real axis until \(s_p\) and then approaches the turning point in a line parallel to the imaginary axis.
If turning points have the same real part it is sufficient to take one integral which encircles all of the turning points.
Using \Eqref{eq:Kint3} we find
\begin{align*}
 \int_{t_j^\pm}^{t_{j+1}^\pm}\omega(t')dt'=0\,.
\end{align*}
This means that only the integral from \(s_p\) to \(t_p^+\) contributes for the Sauter pulse, since the contributions of the other ones vanish due to the periodic form of \(\omega(t) \).
Accordingly the semiclassical momentum spectrum defined in \Eqref{eq:MomentumSpectrum} takes the form
\begin{align*}
 W^{\text{SC}}_s(\vec{\cm})=\exp\left(-\frac{\pi}{\epsilon}\frac{1}{\gamma^2}\left( \frac{\gamma \cm_+}{m}+\frac{ \gamma  \cm_-}{m}-2\right)\right)\,.
\end{align*}
This can be compared to the exact result, which for instance can be obtained in the real-time DHW formalism and is \cite{Hebenstreit2010,Hebenstreit2011}
\begin{widetext}
\begin{align*}
 W_s(\vec{\cm})=\frac{\sinh\left(\frac{1}{2}\frac{\pi}{\epsilon}\frac{1}{\gamma^2}\left[2+\frac{\gamma \cm_+}{m}-\frac{\gamma \cm_-}{m}\right]\right)\sinh\left(\frac{1}{2}\frac{\pi}{\epsilon}\frac{1}{\gamma^2}\left[2 -\frac{\gamma \cm_+}{m }+\frac{\gamma \cm_-}{m }\right]\right)}{\sinh\left(\frac{\pi}{\epsilon}\frac{1}{\gamma} \frac{\cm_+}{m }\right)\sinh\left(\frac{\pi}{\epsilon}\frac{1}{\gamma} \frac{\cm_+}{m }\right)}\,.
\end{align*}
\end{widetext}
Using the fact that
 \( \sinh(x)\approx2 \exp(x) \)
for large $x$ we find for \(\epsilon\gamma\sim \nicefrac1{\tau m}\ll1\)
\begin{align*}
 W_s(\vec{\cm})\overset{\tau m\gg1}\approx W^{\text{SC}}_s(\vec{\cm})
\end{align*}
such that the semiclassical result is approximating the exact one well for long enough pulses.
As described in section \ref{sec:cmp} for shorter pulses the turning points get too close in the complex plane and the approximation in \Eqref{eq:APPROX} breaks down.
\clearpage
%\bibliography{Bib}

%

%-------------------------------------------------------------------------------
\end{document}